\documentclass[aps,pra,twocolumn]{revtex4}
\usepackage{graphicx}
\usepackage{dcolumn}
\usepackage{amsmath}
\usepackage{amssymb}
\usepackage{float}
\usepackage{epstopdf}

\def\itOm{{\mathit{\Omega}}}

\def\xc{_{\rm xc}}

\def\eps{{\epsilon}}
\def\om{{\omega}}
\def\rv{{\bf r}}
\def\fv{{\bf f}}
\def\sv{{\bf s}}

\def\Rv{{\bf R}}

\def\Fv{\underline{f}}
\def\Rv{\underline{r}}
\def\Qv{\underline{q}}
\def\Uv{\underline{u}}
\def\Xv{\underline{x}}

\def\beq{\begin{equation}}
\def\eeq{\end{equation}}

\def\tPs{\widetilde{\Psi}}
\def\bPs{\overline{\Psi}}

\def\ql{\underline{q}}
\def\ul{\underline{u}}
\def\el{\underline{e}}
\begin{document}
\title{Electronic zero-point oscillations in the strong-interaction limit \\ of density functional theory}
\author{Paola Gori-Giorgi,$^{1,2}$ Giovanni Vignale,$^3$ and Michael Seidl$^4$}
\affiliation{$^1$Laboratoire de Chimie Th\'{e}orique, CNRS,
Universit\'{e} Pierre et Marie Curie, 4 Place Jussieu, 75252 Paris, France \\
$^2$Afdeling Theoretische Chemie, Vrije Universiteit, De Boelelaan 1083, 1081	HV Amsterdam, The Netherlands\\
$^3$Department of Physics and Astronomy,
University of Missouri, Columbia, Missouri 65211, USA \\
$^4$Institute of Theoretical Physics,
University of Regensburg, 93040 Regensburg, Germany}
\date{\today}

\begin{abstract}
The exchange-correlation energy in Kohn-Sham density functional theory can be expressed exactly in terms of the change in the expectation of the electron-electron repulsion operator when, in the many-electron hamiltonian, this same operator is multiplied by a real parameter $\lambda$ varying between 0 (Kohn-Sham system) and 1 (physical system). In this process, usually called adiabatic connection, the one-electron density is kept fixed by a suitable local one-body potential. The strong-interaction limit of density functional theory, defined as the limit $\lambda\to\infty$, 
turns out to be, like the opposite non-interacting Kohn-Sham limit ($\lambda\to 0$) mathematically simpler than the physical ($\lambda=1$) case, and can be used to build an approximate interpolation formula between $\lambda\to 0$ and $\lambda\to\infty$ for the exchange-correlation energy. Here we extend the exact treatment of the $\lambda\to\infty$ limit [Phys. Rev. A {\bf 75}, 042511 (2007)] to the next leading term, describing zero-point oscillations of strictly correlated electrons, with numerical examples for small spherical atoms. We also propose an improved approximate functional for the zero-point term and a revised interpolation formula for the exchange-correlation energy satisfying more exact constraints.  
\end{abstract}

\maketitle
\section{Introduction}
Kohn-Sham (KS) density functional theory (DFT) \cite{HohKoh-PR-64,KohSha-PR-65,Koh-RMP-99} is a very successful method for electronic structure calculations thanks to its unique combination of low computational cost and remarkable accuracy. 
In the Kohn-Sham formalism, the ground-state energy of a many-electron system in a given external potential $\hat{V}_{\rm ext}=\sum_{i=1}^N v_{\rm ext}(\rv_i)$ is rewritten as a functional of the one-electron density $\rho(\rv)$,
\beq
E[\rho]=F[\rho]+\int d^3r\, \rho(\rv) v_{\rm ext}(\rv),
\eeq
where
\beq
F[\rho]=\min_{\Psi\to\rho}\langle \Psi|\hat{T}+\hat{V}_{\rm ee}|\Psi \rangle, 
\label{eq_F}
\eeq
with the operators (in Hartree atomic units $e=m=\hbar=a_0=1$ used throughout)
\begin{eqnarray}
	\hat{T} & = & -\frac{1}{2}\sum_{i=1}^N\nabla_i^2 \\
\hat{V}_{\rm ee} & = & \frac12\sum_{i,j=1}^N\frac{1-\delta_{ij}}{|\rv_i-\rv_j|}.
\end{eqnarray}
In Eq.~(\ref{eq_F}) the minimum search is carried over all antisymmetric wavefunctions yielding a given density $\rho$ \cite{Lev-PNAS-79}. The universal functional $F[\rho]$ of Eq.~(\ref{eq_F}) is further divided into 
\beq
F[\rho]=T_s[\rho]+U[\rho]+E_{\rm xc}[\rho],
\label{eq_splitF}
\eeq
where the non-interacting kinetic energy functional $T_s[\rho]$ is obtained by replacing 
$\hat{V}_{\rm ee}$ with zero in Eq.~(\ref{eq_F}),
\beq
T_s[\rho]=\min_{\Psi\to\rho}\langle \Psi|\hat{T}|\Psi\rangle,
\eeq
and the Hartree functional $U[\rho]$ is the classical electrostatic repulsion energy
\beq
U[\rho]=\frac12\int d^3r\int d^3r'\frac{\rho(\rv)\rho(\rv')}{|\rv-\rv'|}.
\eeq
The only quantity that needs to be approximated is the functional for the
exchange-correlation energy, $E_{\rm xc}[\rho]$, defined as the quantity needed to make
Eq.~(\ref{eq_splitF}) exact. The great success of KS DFT in solid state physics stems from the fact that even the simplest approximation for $E_{\rm xc}[\rho]$, the local-density approximation (LDA), already gives remarkable results for basic properties of simple solids. A fundamental step forward to improve the solid-state physics results,  and to spread the use of KS DFT into the quantum chemistry world, has been the advent of generalized gradient approximations (GGA), which are, to a large amount, due to the work of John P. Perdew and his coworkers \cite{Per-INC-91,PerCheVosJacPedSinFio-PRB-92,PerBurErn-PRL-96}. 
%However, these limited-form
%{\em explicit} density functionals have reached a limit of accuracy where improvement for some systems or properties is often associated to a worsening elsewhere. 

Despite its success in scientific areas now ranging from material science to biology, KS-DFT is far from being perfect, and a huge effort is put nowadays in trying to improve the approximations for $E\xc[\rho]$ (for recent reviews see, e.g., \cite{Mat-SCI-02,PerRuzTaoStaScuCso-JCP-05}). The focus of a large part of the scientific community working in this area has shifted from seeking explicit functionals of the density (like the GGA's), to implicit density functionals that 
construct the exchange-correlation energy from the KS orbitals. For example, predicted atomization energies of molecules have been improved by meta-GGA's (MGGA) \cite{PerKurZupBla-PRL-99,TaoPerStaScu-PRL-03} which make use of the orbital kinetic energy density, by hybrid functionals (see, e.g., \cite{Bec-JCP-93a,Bec-JCP-93}) which mix a fraction of exact exchange with GGA exchange and correlation, and by range-separated hybrids, in which only exact long- or short-range exact exchange is used (see, e.g.,\cite{IikTsuYanHir-JCP-01,HeyScuErn-JCP-03,VydHeyKruScu-JCP-06,VydScu-JCP-06,HenIzmScuSav-JCP-07}).

The next step \cite{PerSmi-INC-01} towards higher accuracy  could be
fully non-local functionals, which use 100\% of exact exchange  (for a recent review, see \cite{KumKro-RMP-08}). Despite several attempts and the increasing understanding of the crucial problems \cite{PerRuzCsoVydScuStaTao-PRA-07}, the construction of a fully non-local correlation energy functional compatible with exact exchange is still an issue. A possible way to address this problem is to use the information contained in the strong-interaction limit of DFT \cite{SeiPerKur-PRL-00}. To explain this strategy,
we have first to recall an exact formula \cite{LanPer-SSC-75} for $E_{\rm xc}[\rho]$,   
\beq
E_{\rm xc}[\rho]=\int_0^1d\lambda\,W_\lambda[\rho].
\label{CCI}
\eeq
The integrand $W_\lambda[\rho]$ is given by
\beq
W_\lambda[\rho]=\langle\Psi_\lambda[\rho]|\hat{V}_{\rm ee}|\Psi_\lambda[\rho]\rangle-U[\rho],
\eeq
where $\Psi_\lambda[\rho]$, for a given value of $\lambda\ge0$, is the wavefunction
that minimizes $\langle \Psi|\hat{T}+\lambda \hat{V}_{\rm ee}|\Psi\rangle$ and yields the density $\rho$. If $\rho$ is $v$-representable for all $\lambda\ge 0$, $\Psi_\lambda[\rho]$ is the ground state of a fictitious $N$-electron system with the Hamiltonian
\beq
\hat{H}_\lambda[\rho]=\hat{T}+\lambda\hat{V}_{\rm ee}+\hat{V}_{\rm ext}^\lambda[\rho],
\label{eq_Hlam}
\eeq
where the $\lambda$-dependent external potential,
\beq
\hat{V}^\lambda_{\rm ext}[\rho]=\sum_{i=1}^Nv^\lambda_{\rm ext}([\rho];\rv_i),
\eeq
ensures that $\hat{H}_\lambda[\rho]$ have the same given ($\lambda=1$) ground-state density
$\rho(\rv)$ for all $\lambda$.  
When $\lambda= 0$, the Hamiltonian of Eq.~(\ref{eq_Hlam}) becomes the KS Hamiltonian, and 
$v^{\lambda=0}_{\rm ext}([\rho];\rv)=v_{\rm KS}(\rv)$, the familiar KS potential, while for $\lambda=1$ we recover the Hamiltonian of the physical system. 

We can use perturbation theory to obtain an expansion of $W_\lambda[\rho]$ in powers of $\lambda$ starting from $\lambda=0$,
\beq
W_\lambda[\rho]=E_{\rm x}[\rho]+2\,\lambda E_{\rm c}^{\rm GL2}[\rho]+O(\lambda^2),
\label{wil}
\eeq
where $E_{\rm x}[\rho]$ is the exchange energy and $E_{\rm c}^{\rm GL2}[\rho]$ is the second-order correlation
energy in G{\"o}rling-Levy \cite{GorLev-PRA-94} perturbation theory. This perturbation series expansion,  however, seems to have a finite radius of convergence
($\lambda_c$) which for many atoms and molecules is less
than 1, $\lambda_c<1$ \cite{SeiPerKur-PRL-00,JenJorAgrOls-JCP-96}. Moreover, evaluating terms of ever higher order becomes impracticably
expensive. Nevertheless, the exact lowest-order terms $E_{\rm x}[\rho]$ and $E_{\rm c}^{\rm GL2}[\rho]$
can be used for an alternative approach \cite{SeiPerKur-PRL-00}, called interaction-strength interpolation (ISI), to approximate the integrand in Eq.~\eqref{CCI}.  The basic idea of ISI is to  combine the $\lambda\to 0$ limit of Eq.~\eqref{wil} with the
information from the opposite strong-interaction limit, $\lambda\to\infty$, to construct an
interpolation formula for $W_\lambda[\rho]$. This way, the information on the physical system at $\lambda=1$ is extracted from an interpolation between $\lambda\to 0$ and $\lambda \to \infty$. 

In the strong-interaction limit, $\lambda\to\infty$, we will show in the next sections that $W_\lambda[\rho]$ has the asymptotic expansion
\beq
W_\lambda[\rho]=W_\infty[\rho]+\frac{W'_\infty[\rho]}{\sqrt{\lambda}}+O(\lambda^{-p}),
%\qquad(\lambda\to\infty),
\label{sil}
\eeq
where $p\ge\frac54$. The expansion \eqref{sil} was justified from physical arguments in 
Refs.~\cite{Sei-PRA-99,SeiPerLev-PRA-99}, and a simple approximation for the two functionals $W_\infty[\rho]$ and $W'_\infty[\rho]$, the point-charge plus continuum (PC) model \cite{SeiPerKur-PRA-00},  has been used for the ISI, yielding atomization energies with errors within 4.3 kcal/mol \cite{SeiPerKur-PRL-00}. In a recent paper \cite{SeiGorSav-PRA-07},
the functional $W_\infty[\rho]$ of Eq.~\eqref{sil} has been constructed exactly. The main object of the present work is the extension of the exact treatment of Ref.~\cite{SeiGorSav-PRA-07} to the next term, $W'_\infty[\rho]$. 

The paper is organized as follows. In the next Sec.~\ref{sec_SCE}, we briefly review the results of Ref.~\cite{SeiGorSav-PRA-07}, recalling that the strong-interaction limit of DFT reduces to a $3N$-dimension classical equilibrium problem whose minimum is degenerate over a three-dimensional subspace. In Secs.~\ref{sec_minimum} and \ref{sec_normmod} we define local curvilinear coordinates based on the local normal modes around the degenerate minimum. These
local curvilinear coordinates will be used,  in Sec.~\ref{sec_Hexp}, to expand the Hamiltonian of Eq.~\eqref{eq_Hlam} for $\lambda\to\infty$, up to the order $\lambda^{1/4}$. The corresponding
expansion of $\Psi_\lambda[\rho]$ is carried out in Sec.~\ref{sec_PsiExp}, and the exact expression for $W_{\infty}'[\rho]$ is obtained in Sec.~\ref{sec_Wprime}, where we also report numerical results for small spherical atoms, and we propose an improved PC functional for $W_{\infty}'[\rho]$. In Sec.~\ref{sec_newISI} we revise the interpolation formula for the ISI functional in order to satisfy the exact expansion of Eq.~\eqref{sil} up to $O(\lambda^{-1})$. The last Sec.~\ref{sec_conc} is devoted to conclusions and perspectives. More details of the derivation of our expansion are given in Appendix~\ref{app_T}, and a fully analytic example is reported in Appendix~\ref{app_ana}.  
 
\section{Strictly correlated electrons (SCE)}
\label{sec_SCE}
In the $\lambda\to\infty$ limit it has been shown \cite{Sei-PRA-99,SeiGorSav-PRA-07} that,  in order to keep the $N$ electrons in the given density $\rho$, the external potential in Eq.~(\ref{eq_Hlam}) must compensate the infinitely strong interelectronic repulsion, thus becoming proportional to $\lambda$,
\beq
\lim_{\lambda\to\infty}\frac{v^\lambda_{\rm ext}([\rho],\rv)}{\lambda}=v_{\rm SCE}([\rho],\rv),
\eeq
with a smooth finite function $v_{\rm SCE}([\rho],\rv)$. (For brevity, the argument $[\rho]$ will be often dropped in the following).

The leading term in Eq.~(\ref{eq_Hlam}) when $\lambda\to\infty$ is then a purely multiplicative
potential-energy operator, 
\beq
\hat{H}_{\lambda\to\infty}[\rho]=\lambda (\hat{V}_{\rm ee}+\hat{V}_{\rm SCE})+O(\sqrt{\lambda}). 
\eeq
The square $|\Psi_{\lambda\to\infty}[\rho]|^2$ of the corresponding wavefunction is a distribution that is zero everywhere except for electronic configurations for which $\hat{V}_{\rm ee}+\hat{V}_{\rm SCE}$ has its global minimum. In order to guarantee a given {\em smooth} density $\rho(\rv)$ in such a ``classical'' state,, this global minimum must be degenerate over a three-dimensional subspace of ${\sf R}^{3N}$ \cite{SeiGorSav-PRA-07} (otherwise, the density would be a sum of delta peaks centered in the equilibrium positions of the $N$ electrons). We call this 
classical state with a smooth density ``strictly correlated electrons'' (SCE). The square of
the SCE wavefunction $|\Psi_{\rm SCE}[\rho]|^2=
|\lim_{\lambda\to\infty}\Psi_\lambda[\rho]|^2$ reads
\begin{eqnarray}
& & |\Psi_{\rm SCE}(\rv_1,...,\rv_N)|^2  =  \frac{1}{N!}\sum_P\int d\sv \frac{\rho(\sv)}{N}
\delta(\rv_1-f_{P(1)}(\sv)) \nonumber \\
& &  \times \delta(\rv_2-f_{P(2)}(\sv))...\delta(\rv_N-\fv_{P(N)}(\sv)),
\label{eq_PsiSCE}
\end{eqnarray}
where $\fv_1,..,\fv_N$ are ``co-motion functions'', with $\fv_1(\rv)=\rv$, and $P$ denotes a permutation of $\{1,...N\}$. This means that the $N$ points $\rv_1,...,\rv_N$ in 3D space found upon simultaneous measurement of the $N$ electronic positions in the SCE state  always obey the
$N-1$ relations
\beq
\rv_i=\fv_i(\rv_1)\qquad(i=2,...,N).
\label{cofs}
\eeq
If the $N-1$ co-motion functions $\fv_i(\sv)$  satisfy the differential equation
\beq
\rho(\fv_i(\rv))d^3 f_i(\rv)=\rho(\rv)d^3 r,
\label{eq_fdiff}
\eeq
together with special transformation properties \cite{SeiGorSav-PRA-07} (see also Ref.~\cite{Liu-THESIS-06}), the SCE wavefunction of
Eq.~(\ref{eq_PsiSCE}) yields the given density $\rho(\rv)$. One has then to find the initial
conditions for the integration of Eqs.~(\ref{eq_fdiff}) that minimize the expectation of 
$\hat{V}_{\rm ee}$. The leading coefficient $W_\infty[\rho]$ in Eq.~\eqref{sil} has a simple analytic expression in terms of
the $\fv_i(\sv)$ [see Eq.~\eqref{eq_Winfty}], and has been evaluated for spherical atoms with up to $N=10$ electrons \cite{SeiGorSav-PRA-07}. 

In order to treat the next leading term, $W_{\infty}'[\rho]$ of Eq.~\eqref{sil}, we have to
consider the next terms in the $\lambda\to\infty$ expansion of the Hamiltonian of Eq.~(\ref{eq_Hlam}), i.e., the kinetic energy $\hat{T}$ and the next orders of $\hat{V}_{\rm ext}^\lambda$. Physically, we expect that $W_{\infty}'[\rho]$ is determined by zero-point oscillations around the degenerate SCE minimum. In the following, we give a formal justification to this physical argument.

\section{The SCE potential-energy minimum}
\label{sec_minimum}
Writing $\Rv\equiv(\rv_1,...,\rv_N)\in{\sf R}^{3N}\equiv\itOm$,
we consider the asymptotic potential-energy function ($\itOm\to{\sf R}$),
\begin{eqnarray}
E_{\rm pot}(\Rv)&:=&\lim_{\lambda\to\infty}\frac{\hat{H}_\lambda[\rho]}{\lambda}
\nonumber\\
                & =&\frac12\sum_{i,j=1}^N\frac{1-\delta_{ij}}{|\rv_i-\rv_j|}
                                        +\sum_{i=1}^Nv_{\rm SCE}(\rv_i)\nonumber\\
                & =&\hat{V}_{\rm ee}+\hat{V}_{\rm SCE}.
\label{Epot}
\end{eqnarray}
As said, the SCE external potential $v_{\rm SCE}(\rv)$ has the very special property that the function
$E_{\rm pot}(\Rv)$ has a degenerate minimum $E_{\rm SCE}$ on the 3D subset
\beq
\itOm_0=\{\Fv(\sv)\,|\,\sv\in{\sf R}^3\}\subset\itOm,
\label{Om0}
\eeq
where $\Fv(\sv)=(\sv,\fv_2(\sv),...,\fv_N(\sv))$, with the ${\sf R}^3\to{\sf R}^3$
co-motion functions $\fv_i(\sv)$. In other words, for all $\Rv\in\itOm_0$,
the function $E_{\rm pot}(\Rv)$ assumes the same constant value
\beq
E_{\rm SCE}=W_\infty[\rho]+U[\rho]+\sum_{i=1}^Nv_{\rm SCE}(\fv_i(\sv))
\label{ESCE}
\eeq
which,
in particular, is its global minimum within $\itOm$. For illustration, see the analytical example of Eq.~\eqref{eq_Epotana} in Appendix~\ref{app_ana}.

In the very limit $\lambda\to\infty$, when $\hat{H}_\lambda[\rho]\to\lambda E_{\rm pot}(\Rv)+O(\sqrt{\lambda})$,
the square of the wave function $|\Psi_\lambda[\rho]|^2$ becomes the distribution $|\Psi_{\rm SCE}[\rho]|^2$ of Eq.~\eqref{eq_PsiSCE},
which is strictly zero almost everywhere in $\itOm$ except for the 3D subset $\itOm_0$
where $E_{\rm pot}(\Rv)$ is minimum \cite{SeiGorSav-PRA-07},
\beq
\Psi_{\rm SCE}([\rho],\Rv)\equiv0\qquad\forall\;\Rv\in \itOm \backslash \itOm_0.
\label{distr}
\eeq
For large, but finite $\lambda\gg1$, the electrons are expected to perform  small zero-point
oscillations about the SCE configurations $\Rv\in\itOm_0$, within a narrow $3N$-D
``envelope'' $\itOm_\eps$ (with a small width $\eps>0$) of the 3D subset $\itOm_0\subset\itOm$,
\beq
\itOm_\eps=\{\Rv\in\itOm\,|\,d(\Rv,\itOm_0)<\eps\}.
\eeq
Here, for a given $\Rv\in\itOm$, the quantity
\beq
d(\Rv,\itOm_0):=\min_{\sv\in{\sf R}^3}|\Rv-\Fv(\sv)|
\label{mindist}
\eeq
is the minimum $3N$-D distance between $\Rv$ and any $\Fv(\sv)\in\itOm_0$.     %for any $\sv\in{\sf R}^3$.
%More generally, for small finite $\eps>0$, let $\itOm_\eps\subset\itOm$ be the $3N$-D (!) 
%"envelope" of the 3D subset $\itOm_0\subset\itOm$ that consists of all the points $\Rv\in\itOm$
%whose $3N$-D distance from $\itOm_0$,
Notice that $\itOm_0\subset\itOm_\eps\subset\itOm$ and $\itOm_0=\lim_{\eps\to0}\itOm_\eps$.

For $\Rv\in\itOm_\eps$, $E_{\rm pot}(\Rv)$ may be expanded about $\Rv(\sv)\in\itOm_0$,
\begin{eqnarray}
E_{\rm pot}(\Rv)&=&E_{\rm SCE}
+\frac12\sum_{\mu,\nu=1}^{3N}M_{\mu\nu}(\sv)\times
                               \nonumber\\
&&\times(r_\mu-f_\mu(\sv))(r_\nu-f_\nu(\sv))+...
\label{Taylor0}
\end{eqnarray}
Since $E_{\rm pot}(\Rv)$ is minimum at $\Rv=\Fv(\sv)$, there are no first-order terms. [The dots
represent the terms of third and higher orders.] For any given $\sv\in{\sf R}^3$, the Hessian matrix
$M_{\mu\nu}(\sv)$ in the second-order term
has $3N$ non-negative eigenvalues $\om_\mu(\sv)^2$ which can be labeled such that
\begin{eqnarray}
\om_\mu(\sv)^2=0&&(\mu=1,2,3),\nonumber\\
\om_\mu(\sv)^2>0&&(\mu=4,...,3N).
\end{eqnarray}
The corresponding $3N$-D normalized
eigenvectors $\el^\mu(\sv)$, with components $e_\sigma^\mu(\sv)$ ($\sigma=1,...,3N$),
are pairwise orthogonal,
\beq
\el^\mu(\sv)\cdot\el^\nu(\sv)\equiv
\sum_{\sigma=1}^{3N}e_\sigma^\mu(\sv)e_\sigma^\nu(\sv) = \delta_{\mu\nu}.
\eeq
The first three eigenvectors, with zero eigenvalues, lie in the space ``tangential'' to $\itOm_0$,
the remaining $3N-3$ eigenvectors are ``orthogonal'' to $\itOm_0$,
\beq
\el^\mu(\sv)\cdot\frac{\partial\Fv(\sv)}{\partial s_\alpha}=0
\qquad(\mu=4,...,3N,\quad\alpha=1,2,3),
\eeq
where $\alpha=1,2,3$ denotes the three cartesian components ($x,y,z$) of $\sv$.

\section{Local normal modes}
\label{sec_normmod}

For sufficiently small $\eps>0$, we use these eigenvectors to introduce a set of $3N$ curvilinear
coordinates in $\itOm_\eps$. 
A given point $\Rv=(r_{11},r_{12},r_{13},...,r_{N1},r_{N2},r_{N3})\in\itOm_\eps$, 
is written in terms of these local curvilinear coordinates as follows. The first three curvilinear coordinates are the cartesian coordinates $s_1,s_2,s_3$ of the minimizing vector
$\sv$ in Eq.~\eqref{mindist}, fixed by the condition that the $3N$-D vector $\Rv-\Fv(\sv)$ in $\itOm$
is orthogonal to $\itOm_0$ in the point $\Fv(\sv)$,
\beq
(\Rv-\Fv(\sv))\cdot\frac{\partial\Fv(\sv)}{\partial s_\alpha}=0\qquad(\alpha=1,2,3).
\eeq
%For a given point $\Rv\in\itOm_\eps$, choose $\sv\in{\sf R}^3$ such that
%the $3N$-D distance $|\Rv-\Rv(\sv)|$ between $\Rv\in\itOm_\eps$ and $\Fv(\sv)\in\itOm_0$ is minimum.
%The components $s_1,s_2,s_3$ of $\sv$ are the first three coordinates of $\Rv\in\itOm_\eps$.
The remaining    $3N-3$ 
coordinates are the projections $q_4,...,q_{3N}$ of $\Rv-\Fv(\sv)$ onto the
local eigenvectors $\el^4(\sv),...,\el^{3N}(\sv)$,
\beq
%q_\mu=(\Rv-\Fv(\sv))\cdot\el^\mu(\sv)\qquad(\mu=4,...,3N).
\Rv-\Fv(\sv)=\sum_{\mu=4}^{3N}q_\mu\,\el^\mu(\sv).
\label{qr}
\eeq
The first three eigenvectors $\el^{1,2,3}(\sv)$ are not needed, since they are tangential to $\itOm_0$
at the point $\Fv(\sv)$ and therefore orthogonal to $\Rv-\Fv(\sv)$. Inverting Eq.~\eqref{qr} yields
\beq
q_\mu=\el^\mu\cdot(\Rv-\Fv(\sv))\qquad(\mu=4,...,3N).
\eeq

%Within the close vicinity of $\Rv(\sv)$ in $\itOm$, these eigenvectors may serve as local new
%coordinate axes in $\itOm$. The resulting

%These coordinates $(s_1,s_2,s_3,q_4,...,q_{3N})$ are related to the original ones,
%$\Rv=(r_{11},...,r_{N3})$, by
%\begin{eqnarray}
%r_{1\alpha}&=&s_\alpha\qquad(\alpha=1,2,3),\nonumber\\
%r_{i\alpha}&=&f_{i\alpha}(\sv)+\sum_{\mu=4}^{3N}w_{i\alpha}^\mu(\sv)\,q_\mu\qquad(i=2,...,N).
%\end{eqnarray}
%The first three terms $\mu=1,2,3$ are excluded from the sum, since for any given point
%$\Rv\in\itOm_\eps$, $\sv\in{\sf R}^3$ is chosen such that $q_1=q_2=q_3=0$.

%The $3N-3$ columns $\widetilde{\el}^4(\sv),...,\widetilde{\el}^{3N}(\sv)$, with
%the $3N-3$ components $w^\mu_{i\alpha}(\sv)$ each ($i=2,...,N;\alpha=1,2,3$),
%form an orthogonal quadratic matrix,
%\beq
%\widetilde{\el}^\mu(\sv)\cdot\widetilde{\el}^\nu(\sv)\equiv
%\sum_{i=2}^N\sum_{\alpha=1}^3e_{i\alpha}^\mu(\sv)e_{i\alpha}^\nu(\sv) = \delta_{\mu\nu}.
%\eeq
%Since the quadratic matrix with the $3N-3$ columns $\el^4(\sv),...,\el^{3N}(\sv)$
%is orthogonal,
%Therefore, the inverse transformation is achieved with the transposed matrix,
%\begin{eqnarray}
%s_\alpha&=&r_{1\alpha}\qquad(\alpha=1,2,3),\nonumber\\
%q_\mu&=&\sum_{i=2}^N\sum_{\alpha=1}^3e_{i\alpha}^\mu(\sv)\,\Big[r_{i\alpha}-f_{i\alpha}(\sv)\Big]
%\;(\mu=4,...,3N).
%\end{eqnarray}

For these new curvilinear coordinates, we also write
\beq
(s_1,s_2,s_3,q_4,...,q_{3N})=(\sv,\ql).
\eeq
Notice that $\Rv$ has $3N$ components, while $\Qv$ has only $3N-3$ ones.
In this notation, Eq.~\eqref{qr} reads
%\beq
%q_\mu=\sum_{\nu=4}^{3N}e_\nu^\mu(\sv)(r_\nu-r_\nu(\sv))\qquad(\mu=4,...,3N).
%\eeq
%Since the $e_\nu^\mu(\sv)$, with $\mu,\nu=4,...,3N$, form an orthogonal $(3N-3)\times(3N-3)$ matrix,
%the inverse transformation is given by
\beq
r_\nu=f_\nu(\sv)+\sum_{\mu=4}^{3N}e_\nu^\mu(\sv)q_\mu\qquad(\nu=1,...,3N).
\label{rq}
\eeq
This is the transformation formula between the cartesian coordinates $\Rv$ and the ``local normal modes'' $(\sv,\Qv)$
in the $3N$-D configuration space $\itOm$. 

In terms of the $q_\mu$, the second-order contribution in the Taylor expansion
\eqref{Taylor0} becomes diagonal,
\begin{eqnarray}
\widetilde{E}_{\rm pot}(\sv,\Qv)&=&E_{\rm SCE}+\frac12\sum_{\mu=4}^{3N}\om_\mu(\sv)^2q_\mu^2+\nonumber\\
&&+\frac1{3!}\sum_{\mu,\nu,\sigma=4}^{3N}E^{(3)}_{\mu\nu\sigma}(\sv)q_\mu q_\nu q_\sigma+...
\label{Taylor}
\end{eqnarray}
Here, the third-order term is derived from the corresponding term in Eq.~\eqref{Taylor0}
(in the present notation),
\begin{eqnarray}
&&\hspace{-10mm}\frac1{3!}\sum_{\xi,\eta,\zeta=1}^{3N}
\frac{\partial^3E_{\rm pot}(\Rv)}{\partial r_\xi\partial r_\eta\partial r_\zeta}\Big|_{\Rv=\Fv(\sv)}
\times\nonumber\\
&&\times(r_\xi-f_\xi(\sv))(r_\eta-f_\eta(\sv))(r_\zeta-f_\zeta(\sv)).
\end{eqnarray}
Using here Eq.~\eqref{rq} for $r_\nu-f_\nu(\sv)$, we find
\beq
E^{(3)}_{\mu\nu\sigma}(\sv)=\sum_{\xi,\eta,\zeta=1}^{3N}
\frac{\partial^3E_{\rm pot}(\Rv)}{\partial r_\xi\partial r_\eta\partial r_\zeta}\Big|_{\Rv=\Fv(\sv)}
e^\mu_\xi(\sv)e^\nu_\eta(\sv)e^\sigma_\zeta(\sv).
\eeq
Substituting Eq.~\eqref{rq} for $\Rv$ in the wave function
$\Psi_\lambda(\Rv)$ that represents the state $\Psi_\lambda[\rho]$ yields the transformed wave function
$\tPs_\lambda(\sv,\Qv)$. While the original wave function obeys
\beq
\int d^3r_1...\int d^3r_N|\Psi_\lambda(\Rv)|^2\equiv\int d\Rv|\Psi_\lambda(\Rv)|^2=1,
\eeq
the transformed one is normalized according to
\beq
\int d^3s\int d\Qv\,J(\sv,\Qv)\,\big|\tPs_\lambda(\sv,\Qv)\big|^2=1
\label{tPnorm}
\eeq
where $J(\sv,\Qv)$ is the Jacobian associated with the coordinate transformation \eqref{rq}, see
Eq.~\eqref{Jac} in Appendix \ref{app_T}.

For sufficiently large $\lambda\gg1$,
the wave function $\tPs_\lambda(\sv,\Qv)$ strongly suppresses all configurations $\Rv\in\itOm$ except
for the ones inside the narrow envelope $\itOm_\eps$ of the 3D subset $\itOm_0$. This means that
$\tPs_\lambda(\sv,\Qv)$ is essentially
different from zero only for $(q_4^2+...+q_{3N}^2)^{1/2}<\eps$, where $\eps$ decreases with growing
$\lambda\gg1$ and goes to zero in the limit $\lambda\to\infty$.

More precisely, since the quadratic term in Eq.~\eqref{Taylor} is multiplied by $\lambda$ in the
Hamiltonian \eqref{eq_Hlam}, the scale of the quantum fluctuation is $\eps\sim\lambda^{-1/4}\equiv\alpha$ for $\lambda\to\infty$.
Therefore, it will be useful to switch for a given value of $\lambda\gg1$ from
the present curvilinear coordinates $(\sv,\ql)$ to scaled coordinates $(\sv,\Uv)$ where
\beq
\Uv=\lambda^{1/4}\Qv\quad\Leftrightarrow\quad\Qv=\alpha\Uv \qquad(\alpha=\lambda^{-1/4}).
\eeq
This second transformation yields the wave function
\beq
\bPs_\alpha(\sv,\Uv)=\tPs_\lambda(\sv,\alpha\Uv).
\eeq
According to Eq.~\eqref{tPnorm}, we now have
\beq
\int d^3s\int d\Uv\,K_\alpha(\sv,\Uv)\,\big|\bPs_\alpha(\sv,\Uv)\big|^2=1,
\label{bPnorm}
\eeq
with the scaled Jacobian
\beq
K_\alpha(\sv,\Uv)=\alpha^{3N-3}J(\sv,\alpha\Uv).
\eeq

Later on, we shall make use of the expansion
\beq
J(\sv,\Qv)=J(\sv,\underline{0})+\sum_{\mu=4}^{3N}J^{(1)}_\mu(\sv)q_\mu+O(q_\nu^2),
\label{Jex}
\eeq
whose derivation is reported in Appendix~\ref{app_T}.

\section{Expansion of the Hamiltonian}
\label{sec_Hexp}

To obtain an expansion for large $\lambda\gg1$ (or, equivalently, for small
$\alpha\equiv\lambda^{-1/4}\ll1)$, we must express the Hamiltonian $\hat{H}_\lambda[\rho]$ of Eq.~\eqref{eq_Hlam} in terms
of the scaled coordinates $(\sv,\Uv)$. To this end, we split $\hat{H}_\lambda[\rho]$ into three pieces,
\beq
\hat{H}_\lambda[\rho]=\hat{T}+\lambda E_{\rm pot}(\Rv)
                             +(\hat{V}_{\rm ext}^\lambda-\lambda\hat{V}_{\rm SCE}),
\label{split}
\eeq
and treat these separately now.   %under (a), (b), and (c) below.

\subsection{Kinetic energy (first term)}

For the kinetic-energy operator $\hat{T}$, the $3N$-D Laplacian is obtained in Appendix A
in terms of the curvilinear coordinates $q_\mu$ from the general transformation rule
\beq
\sum_{i=1}^3\nabla_i^2\equiv\sum_{\mu=1}^{3N}\frac{\partial^2}{\partial r_\mu^2}
=\sum_{\mu,\nu=1}^{3N}\frac1{\sqrt{G}}\frac{\partial}{\partial q_\mu}
                        \Big(\sqrt{G}\,G^{\mu\nu}\frac{\partial}{\partial q_\nu}\Big).
\label{eq_lapl}
\eeq
(To simplify the notation, we write $s_\mu\equiv q_\mu$ for $\mu=1,2,3$ in this subsection.)
Here, the matrix $G^{\mu\nu}$ is the inverse of the metric tensor $G_{\mu\nu}$, defined by
\beq
G_{\mu\nu}=\sum_{\xi=1}^{3N}\frac{\partial r_\xi}{\partial q_\mu}\frac{\partial r_\xi}{\partial q_\nu}
\equiv\frac{\partial\Rv}{\partial q_\mu}\cdot\frac{\partial\Rv}{\partial q_\nu},
\eeq
and $G$ is its determinant, $G=\det(G_{\mu\nu})$. Switching in a second step from the $q_\mu$ to the
scaled coordinates $u_\mu$ yields the expansion (see Appendix~\ref{app_T})
\beq
\hat{T}=\sqrt{\lambda}\Big[\hat{T}^{(0)}+\alpha\hat{T}^{(1)}+\alpha^2\hat{T}^{(2)}+O(\alpha^3)\Big].
\label{Texp}
\eeq
The operators $\hat{T}^{(n)}$ are independent of $\lambda$ (or $\alpha\equiv\lambda^{-1/4}$),
\begin{eqnarray}
\hat{T}^{(0)}& = & -\frac12\sum_{\mu=4}^{3N}\frac{\partial^2}{\partial u_\mu^2}, 
\label{eq_T1} \\
\hat{T}^{(1)}& = &-\frac{1}{2}\sum_{\mu=4}^{3N} X_\mu(\sv)\frac{\partial}{\partial u_\mu},  
\label{eq_T2}
\end{eqnarray}
where $X_\mu(\sv)$ is reported in Appendix~\ref{app_T}.
Notice that the $\alpha^2$ term is constant, since $\alpha^2\sqrt{\lambda}=1$.

\subsection{SCE potential energy (second term)}

For the second term in Eq.~\eqref{split}, we use the Taylor expansion \eqref{Taylor}, with $q_\mu=\alpha u_\mu$,
to find
\begin{eqnarray}
&&\hspace*{-6mm}\lambda E_{\rm pot}(\Rv)=\lambda\Big[E_{\rm SCE}
                                 +\frac{\alpha^2}2\sum_{\mu=4}^{3N}\omega_\mu(\sv)^2u_\mu^2\nonumber\\
&&+\frac{\alpha^3}{3!}\sum_{\mu,\nu,\sigma=4}^{3N}
                            E^{(3)}_{\mu\nu\sigma}(\sv)u_\mu u_\nu u_\sigma\nonumber\\
&&+\frac{\alpha^4}{4!}\sum_{\mu,\nu,\sigma,\tau=4}^{3N}
                            E^{(4)}_{\mu\nu\sigma\tau}(\sv)u_\mu u_\nu u_\sigma u_\tau+O(\alpha^5)\Big].
\label{Epotexp}
\end{eqnarray}

\subsection{The remaining external potential (third term)}

For the last term in Eq.~\eqref{split}, we make an ansatz that will later on turn out to be consistent,
\begin{eqnarray}
\hat{V}_{\rm ext}^\lambda-\lambda\hat{V}_{\rm SCE}=
%&\equiv&\sum_{i=1}^N\Big[v_{\rm ext}^\lambda(\rv_i)-v_{\rm SCE}(\rv_i)\Big]\nonumber\\
%&&\hspace*{-24mm}=\sqrt{\lambda}\sum_{i=1}^N\sum_{n=0}^\infty\alpha^nv^{(n)}(\rv_i)\equiv
                       \sqrt{\lambda}\sum_{n=0}^\infty\alpha^nV^{(n)}(\Rv).
\label{ansatz}
\end{eqnarray}
Using Eq.~\eqref{qr} for $\Rv$ and $q_\mu=\alpha u_\mu$, we may expand
%\beq
%V^{(n)}(\Rv)=V^{(n)}\Big(\Fv(\sv)+\alpha\sum_{\mu=4}^{3N}\el^\mu(\sv)u_\mu\Big).
%\eeq
\begin{eqnarray}
V^{(n)}(\Rv)&\equiv&V^{(n)}\Big(\Fv(\sv)+\alpha\sum_{\mu=4}^{3N}\el^\mu(\sv)u_\mu\Big)\nonumber\\
&=&V^{(n)}(\Fv(\sv))
    +\alpha\sum_{\sigma=1}^{3N}V^{(n)}_\sigma(\Fv(\sv))\sum_{\mu=4}^{3N}e_\sigma^\mu(\sv)u_\mu+\nonumber\\
 &&\hspace*{-2mm}+\frac{\alpha^2}2\sum_{\sigma,\tau=1}^{3N}V^{(n)}_{\sigma\tau}(\Fv(\sv))
                            \sum_{\mu,\nu=4}^{3N}e_\sigma^\mu(\sv)e_\tau^\nu(\sv)u_\mu u_\nu+\nonumber\\
 &&+O(\alpha^3).
\end{eqnarray}
Here, the coefficients $V^{(n)}_\sigma$, $V^{(n)}_{\sigma\tau}$, etc.~denote the partial derivatives
of $V^{(n)}(\Rv)$ at $\Rv=\Fv(\sv)$,
\beq
V^{(n)}_{\sigma\tau}(\Fv(\sv))=\frac{\partial^2V^{(n)}(\Rv)}{\partial r_\sigma\partial r_\tau}
\Big|_{\Rv=\Fv(\sv)}\quad\mbox{etc.}
\eeq
Now, Eq.~\eqref{ansatz} yields the expansion
\beq
\hat{V}_{\rm ext}^\lambda-\lambda\hat{V}_{\rm SCE}=
%\sqrt{\lambda}\sum_{n=0}^\infty\alpha^n\sum_{i=1}^Nv^{(n)}(\rv_i)
\sqrt{\lambda}\Big[\hat{V}^{(0)}+\alpha\hat{V}^{(1)}+\alpha^2\hat{V}^{(2)}+O(\alpha^3)\Big],
\label{Vexp}
\eeq
with $\alpha$-independent (multiplicative) operators
\begin{eqnarray}
\hat{V}^{(0)}&=&V^{(0)}(\Fv(\sv)),\nonumber\\
\hat{V}^{(1)}&=&V^{(1)}(\Fv(\sv))
            +\sum_{\sigma=1}^{3N}V^{(0)}_\sigma(\Fv(\sv))\sum_{\mu=4}^{3N}e_\sigma^\mu(\sv)u_\mu,\nonumber\\
\hat{V}^{(2)}&=&V^{(2)}(\Fv(\sv))
            +\sum_{\sigma=1}^{3N}V^{(1)}_\sigma(\Fv(\sv))\sum_{\mu=4}^{3N}e_\sigma^\mu(\sv)u_\mu+\nonumber\\
&&\hspace*{-6mm}+\frac12\sum_{\sigma,\tau=1}^{3N}V^{(0)}_{\sigma\tau}(\Fv(\sv))
                                     \sum_{\mu,\nu=4}^{3N}e_\sigma^\mu(\sv)e_\tau^\nu(\sv)u_\mu u_\nu.
\end{eqnarray}

\subsection{Full Hamiltonian}

Eventually, combining Eqs.~\eqref{Texp}, \eqref{Epotexp}, and \eqref{Vexp}, we obtain the expansion (recall that $\alpha=\lambda^{-1/4}$)
\begin{eqnarray}
\hat{H}_\lambda[\rho]&=&\lambda E_{\rm SCE}\nonumber\\
&+&\sqrt{\lambda}\Big[\hat{H}^{(0)}+\alpha\hat{H}^{(1)}+\alpha^2\hat{H}^{(2)}+O(\alpha^3)\Big]
\label{Hexpd}
\end{eqnarray}
with $\alpha$-independent operators $\hat{H}^{(n)}$. The first two terms read 
\begin{eqnarray}
 	\hat{H}^{(0)} & =& -\frac12\sum_{\mu=4}^{3N}\frac{\partial^2}{\partial u_\mu^2}+V^{(0)}(\Fv(\sv))
	     +\frac12\sum_{\mu=4}^{3N}\omega_\mu(\sv)^2u_\mu^2, \label{eq_H0} \\
 \hat{H}^{(1)}& =& -\frac{1}{2}\sum_{\mu=4}^{3N} X_\mu(\sv)\frac{\partial}{\partial u_\mu}+		V^{(1)}(\Fv(\sv)) +\nonumber \\
 		         &    + & \sum_{\sigma=1}^{3N}V^{(0)}_\sigma(\Fv(\sv))\sum_{\mu=4}^{3N}e_\sigma^\mu(\sv)u_\mu +\nonumber \\
   &    + &  \frac1{3!}\sum_{\mu,\nu,\sigma=4}^{3N}E^{(3)}_{\mu\nu\sigma}(\sv)u_\mu u_\nu u_\sigma. \label{eq_H1}
\end{eqnarray}
%etc.

\section{Expansion of the ground state}
\label{sec_PsiExp}

Due to Eq.~\eqref{Hexpd}, the lowest eigenvalue $E_\lambda[\rho]$ of $\hat{H}_\lambda[\rho]$ (i.~e., its
ground-state energy) has the expansion
\begin{eqnarray}
E_\lambda[\rho]&=&\lambda E_{\rm SCE}+\nonumber\\
&+&\sqrt{\lambda}\Big[E^{(0)}+\alpha E^{(1)}+\alpha^2E^{(2)}+O(\alpha^3)\Big].
\label{Eexpd}
\end{eqnarray}
We define $E'_\alpha[\rho]=E^{(0)}+\alpha E^{(1)}+\alpha^2E^{(2)}+O(\alpha^3)$ as the lowest eigenvalue of the operator
\beq
\hat{H}'_\alpha[\rho]=\hat{H}^{(0)}+\alpha\hat{H}^{(1)}+\alpha^2\hat{H}^{(2)}+O(\alpha^3).
\label{eq_Hprime}
\eeq
 Since $E_{\rm SCE}$ is a constant, $\hat{H}_\lambda[\rho]$ and $\hat{H}'_\alpha[\rho]$,
with $\alpha=\lambda^{-1/4}$, have the same ground state    %which has the expansion
\beq
\bPs_\alpha(\sv,\Uv)=\frac{\Psi^{(0)}+\alpha\Psi^{(1)}+\alpha^2\Psi^{(2)}+O(\alpha^3)}{\sqrt{{\cal N}_\alpha}}.
\label{bPex}
\eeq
For the $\alpha$-dependent normalization constant,
\beq
{\cal N}_\alpha=\int d^3s\int d\Uv\,K_\alpha(\sv,\Uv)\Big|\Psi^{(0)}(\sv,\Uv)+O(\alpha)\Big|^2,
\eeq
we obtain
\beq
{\cal N}_\alpha=\alpha^{3N-3}[1+O(\alpha)]
\label{Nex}
\eeq
when $\Psi^{(0)}$ is normalized according to
\beq
\int d^3s\int d\Uv\,J(\sv,\underline{0})\Big|\Psi^{(0)}(\sv,\Uv)\Big|^2=1.
\eeq

Collecting terms of equal orders $O(\alpha^n)$ in the eigenvalue equation
$\hat{H}'_\alpha[\rho]\bPs_\alpha=E'_\alpha[\rho]\bPs_\alpha$
yields a hierarchy of equations. The leading one is $\hat{H}^{(0)}\Psi^{(0)}=E^{(0)}\Psi^{(0)}$,
where $\hat{H}^{(0)}$ is given by Eq.~\eqref{eq_H0}.
For a given fixed $\sv\in{\sf R}^3$, the Hamiltonian $\hat{H}^{(0)}$ describes an uncoupled set of $3N-3$ harmonic oscillators in 1D. To be more precise, these oscillators are coupled via the  dynamical variable $\sv$, but the dynamics of $\sv$ is much slower, only appearing at orders $O(\lambda^0)$. 
Consequently, the leading term in the wave function factorizes into a product of Gaussians
$\Phi_\omega(u)=(\frac{\omega}{\pi})^{1/4}\,e^{-\omega u^2/2}$, with $\int_{-\infty}^{\infty}du|\Phi_\omega(u)|^2=1$,
\beq
\Psi^{(0)}(\sv,\ul)=C^{(0)}(\sv)\prod_{\mu=4}^{3N}\Phi_{\omega_\mu(\sv)}(u_\mu).
\label{PsiOsc}
\eeq
Since $V^{(0)}(\Fv(\sv))$ is a pure multiplicative operator,
the resulting eigenvalue of $\hat{H}^{(0)}$ is
\beq
E^{(0)}=V^{(0)}(\Fv(\sv))+\sum_{\mu=4}^{3N}\frac{\omega_\mu(\sv)}2.
\label{E0}
\eeq
Due to Eq.~\eqref{Eexpd}, this expression cannot depend on the variable $\sv$, implying a condition
on the $n=0$ coefficient $V^{(0)}(\Rv)$ in our ansatz \eqref{ansatz},
\beq
V^{(0)}(\Fv(\sv))=-\sum_{\mu=4}^{3N}\frac{\omega_\mu(\sv)}2+\mbox{const}\qquad\forall\sv\in{\sf R}^3.
\label{eq_V0}
\eeq
In particular, we have
\beq
E^{(0)}=\int d^3s \frac{\rho(\sv)}{N}\Big[V^{(0)}(\underline{f}(\sv))
+\sum_{\mu=4}^{3N}\frac{\omega_\mu(\sv)}2\Big].
\eeq
The role of the external potential at the order $\sqrt{\lambda}$ in Eq.~(\ref{eq_Hlam}) is thus to keep the degeneracy of the SCE minimum (found at the order $\lambda$) through the order $\sqrt{\lambda}$. This is necessary in order to keep the given smooth density $\rho(\rv)$: if one of the SCE configurations (i.e., a given particular $\sv_0$) had a lower energy than the others, the SCE wavefunction would collapse in that particular $\sv_0$, and the density would become a sum of delta peaks centered in $\fv_i(\sv_0)$ (with $i=1,...,N$).
%Since the probability distribution $\int d\Uv|\Psi^{(0)}(\sv,\ul)|^2=|C^{(0)}(\sv)|^2$ of the variable $\sv$
%must be $\frac1{N}\rho(\sv)$, proportional to the electron density $\rho$, we conclude
%\beq
%C^{(0)}(\sv)=\sqrt{\frac{\rho(\sv)}{N}}.
%\eeq
%
%Transforming the wave function \eqref{PsiOsc} back to the non-scaled coordinates $(\sv,\Qv)=(\sv,\alpha\Uv)
%\Leftrightarrow(\sv,\Uv)=(\sv,\lambda^{1/4}\Qv)$, we see explicitly how the distribution
%$|\Psi_{\rm SCE}|^2$ in Eq.~\eqref{distr} emerges in the limit $\lambda\to\infty$,
%\beq
%\Big|\Psi_{\rm SCE}(\sv,\ql)\Big|^2=
%\lim_{\lambda\to\infty}\lambda^{1/4}\Big|\Psi^{(0)}(\sv,\lambda^{1/4}\ql)\Big|^2.
%\eeq

In order to determine the prefactor $C^{(0)}(\sv)$ of the wave function \eqref{PsiOsc} we observe that in
the wave function $\tPs_\lambda(\sv,\Qv)$, the coordinate $\sv\in{\sf R}^3$ has the probability distribution
\begin{eqnarray}
\rho_\lambda(\sv)&=&\int d\Qv\,J(\sv,\Qv)|\tPs_\lambda(\sv,\Qv)|^2\nonumber\\
                 &=&\int d\Qv\,J(\sv,\Qv)|\bPs_\alpha(\sv,\lambda^{1/4}\Qv)|^2
\label{rholam}
\end{eqnarray}
where $\alpha=\lambda^{-1/4}$. Using Eqs.~\eqref{bPex} and \eqref{Nex}, we find
%\begin{eqnarray}
%\rho_\lambda(\sv)&=&\int d\Qv\,\frac{J(\sv,\Qv)}{\alpha^{3N-3}}
%                      \Big[|\Psi^{(0)}(\sv,\lambda^{1/4}\Qv)|^2+O(\alpha)\Big]\nonumber\\
%                 &=&|C^{(0)}(\sv)|^2\int d\Qv\,J(\sv,\Qv)
%                      \prod_{\mu=4}^{3N}\lambda^{1/4}|\Phi_{\omega_\mu(\sv)(\lambda^{1/4}q_\mu)}|^2
%\label{rholam2}
%\end{eqnarray}
\beq
\rho_\lambda(\sv)=\int d\Qv\,J(\sv,\Qv)\,
                  \frac{\,|\Psi^{(0)}(\sv,\lambda^{1/4}\Qv)|^2}{\alpha^{3N-3}}\,\Big[1+O(\alpha)\Big].
\label{rholam2}
\eeq
In the limit $\lambda\to\infty$ when $\rho_\lambda(\sv)$ must become rigorously
proportional to the electron density $\rho(\sv)$,
\beq
\lim_{\lambda\to\infty}\rho_\lambda(\sv)=\frac{\rho(\sv)}N,
\eeq
the terms $O(\alpha)$ in Eq.~\eqref{rholam2} can be dropped and Eq.~\eqref{PsiOsc} yields
\begin{eqnarray}
\frac{\rho(\sv)}N&=&\lim_{\lambda\to\infty}|C^{(0)}(\sv)|^2\int d\Qv\,J(\sv,\Qv)\times\nonumber\\
&&\times\prod_{\mu=4}^{3N}\lambda^{1/4}|\Phi_{\omega_\mu(\sv)}(\lambda^{1/4}q_\mu)|^2.
\label{eq_limPsi}
\end{eqnarray}
Since $\Phi_\omega(u)$ is a normalized Gaussian, the $\mu$-th factor of the product in Eq.~\eqref{eq_limPsi} approaches
the $\delta$-function $\delta(q_\mu)$ as $\lambda\to\infty$. Therefore, the right-hand side of Eq.~\eqref{eq_limPsi} equals
$|C^{(0)}(\sv)|^2J(\sv,\underline{0})$, implying the result
\beq
|C^{(0)}(\sv)|^2=\frac1N\frac{\rho(\sv)}{J(\sv,\underline{0})}.
\label{C0}
\eeq

The next order in the perturbative treatment of the ground-state energy of Eq.~\eqref{eq_Hprime} leads to
\beq
E^{(1)}=\langle\Psi^{(0)}|\hat{H}^{(1)}|\Psi^{(0)}\rangle=V^{(1)}(\Fv(\sv)).
\label{eq_E1}
\eeq
The same argument used for Eq.~\eqref{eq_V0} yields
\beq
V^{(1)}(\Fv(\sv))={\rm const.},
\eeq
independent on $\sv$. The important point here is that the terms coming from $\hat{T}$ and
$\hat{V}_{\rm ee}$ in the Hamiltonian $\hat{H}^{(1)}$ of Eq.~\eqref{eq_H1} have zero expectation on the ground-state of the harmonic oscillator, so that there is no contribution to this order to the large-$\lambda$ expansion of $W_{\lambda}[\rho]$. As we shall see in the next Sec.~\ref{sec_Wprime}, the order $\sqrt{\lambda}\alpha =\lambda^{1/4}$ in $E_\lambda[\rho]$  of Eq.~(\ref{Eexpd}) corresponds to the order $\lambda^{-3/4}$ in the large-$\lambda$ expansion of $W_{\lambda}[\rho]$.

Notice that, in our treatment of the strong-interaction limit of DFT, we did not consider the effect on the energy of the spin state or, more generally, of the statistics. This is because the electrons are always localized in different regions of space well separated from each other. The effect on the energy of the spin state or of statistics in the $\lambda\to\infty$ limit can be estimated as being of the order $O(e^{-\lambda^{1/4}})$, which is the order of magnitude of the overlap between two different gaussians of Eq.~(\ref{PsiOsc}). 

\section{The coefficient $W'_\infty[\rho]$}
\label{sec_Wprime}
From the expansion of $E_\lambda[\rho]$ of the previous Sec.~\ref{sec_PsiExp}, we can easily compute $W_\lambda[\rho]$ using the Hellmann-Feynmann theorem:
\beq
W_\lambda[\rho]+U[\rho]=\frac{\partial E_\lambda[\rho]}{\partial \lambda}-\int \rho(\rv) \frac{\partial v_{\rm ext}^\lambda(\rv)}{\partial\lambda}d^3 r.
\label{eq_HF}
\eeq
From Sec.~\ref{sec_PsiExp}, we obtain, in the $\lambda\to\infty$ limit,
\begin{eqnarray}
& & E_{\lambda}[\rho]-\int \rho(\rv) v_{\rm ext}^\lambda(\rv)d^3 r  =  \lambda \langle\Psi_{\rm SCE}|\hat{V}_{ee}|\Psi_{\rm SCE}\rangle+ \nonumber \\
& & +  \sqrt{\lambda}\int d^3s\frac{\rho(\sv)}{N}\sum_{\mu=4}^{3N}\frac{\omega_\mu(\sv)}2+O(\lambda^0)
\end{eqnarray}
By differentiating both sides with respect to $\lambda$, from Eq.~\eqref{eq_HF} we obtain
the expansion for $W_\lambda[\rho]$ of Eq.~\eqref{sil} with
\beq
W_\infty[\rho]=\int d^3s \frac{\rho(\sv)}{N}\sum_{i=1}^N\sum_{j>i}^N\frac{1}{|\fv_i(\sv)-\fv_j(\sv)|}-U[\rho],
\label{eq_Winfty}
\eeq
in agreement with the results of Ref.~\cite{SeiGorSav-PRA-07}, and the exact expression for the next leading term,
\beq
W'_\infty[\rho]=\frac12\int d^3s\frac{\rho(\sv)}{N}\sum_{\mu=4}^{3N}\frac{\omega_\mu(\sv)}2.
\label{eq_wprime}
\eeq
This result generalizes (and proves) Eq.~(35) of Ref.~\cite{Sei-PRA-99} for spherical two-electron
densities. 
As shown by Eq.~\eqref{eq_E1}, there is no $\lambda^{-3/4}$ term in $W_{\lambda\to\infty}[\rho]$.
There is also no term $\propto \lambda^{-1}$, which would imply a term $\propto {\rm log}(\lambda)$ in $E_{\lambda}[\rho]$ and thus in the kinetic energy $\langle\Psi_\lambda|\hat{T}|\Psi_\lambda\rangle$. Such a term would violate the known high-density scaling of $\langle\Psi_\lambda|\hat{T}|\Psi_\lambda\rangle$ \cite{LevPer-PRA-85} (see also the erratum of Ref.~\cite{SeiPerKur-PRA-00}).

As an example of application, we have computed $W_{\infty}'[\rho]$ for the same set of spherical (or sphericalized) atomic densities used in Ref.~\cite{SeiGorSav-PRA-07} to evaluate $W_\infty[\rho]$. For each point $(\fv_1(\sv),...,\fv_N(\sv))$ on the degenerate SCE minimum constructed in Ref.~\cite{SeiGorSav-PRA-07}, we have evaluated the hessian matrix, the corresponding eigenvalues $\omega_\mu^2(\sv)$, and thus $W_{\infty}'[\rho]$ of Eq.~\eqref{eq_wprime}. In Table~\ref{tab_wprime} we compare our
results with the approximate PC functional \cite{SeiPerKur-PRA-00},
\beq
W_\infty^{'\rm PC}[\rho]=\int d^3 r\left[C\rho(\rv)^{3/2}+D\frac{|\nabla\rho(\rv)|^2}{\rho(\rv)^{7/6}}\right],
\label{eq_WPC}
\eeq
where $C=1.535$, $D=-0.02558$. 

As explained in Ref.~\cite{SeiGorSav-PRA-07}, the SCE minimum for spherical densities is constructed from a set of radial co-motion functions and the angular minimization is done numerically. When one of the electrons is close to the nucleus, the numerical minimization displays instabilities in the smallest (but non-zero) eigenvalues of the hessian. However, as shown by Eq.~\eqref{eq_wprime}, such configurations are weighted by the density (in the spherically symmetric case by $4\pi \,s^2\rho(s)$) so that the error they introduce is relatively small. This error, however, increases with the number of electrons. The number of digits in our results of Table~\ref{tab_wprime} is determined by this numerical error.  Notice, however, that Table~\ref{tab_wprime} shows that the discrepancy of the PC model with respect to our results is much larger than our estimated numerical errors on the SCE values.

%%%%%%%%%%%%%%%%%%%%%%%%%%%%
\begin{table}
\begin{tabular}{lllc}
\hline \hline
& SCE (H) & PC (H) & error (mH)\\
\hline 
He    &    $0.62084$     &        $0.729$        &           $108$  \\
Li     &   $1.38$      &              $1.622$       &            $240$ \\
Be     &  $2.59$       &             $2.928$       &            $334$ \\
B      &   $4.2$       &               $4.702$    &               $502$ \\
C      &   $6.3$       &               $7.089$    &               $840$ \\
Ne     &   $22$        &              $24.423$    &             $2423$ \\
\hline \hline
\end{tabular}
\caption{Comparison of the values $W'_\infty[\rho]$ in Hartree atomic
units obtained with the SCE construction, and with the PC model \cite{SeiPerKur-PRA-00}. 
The absolute errors of the PC model are also reported.}
\label{tab_wprime}
\end{table}
%%%%%%%%%%%%%%%%%%%%%%%%%%%%%

While the PC model for the coefficient $W_\infty[\rho]$ makes errors of the order of 60~mH \cite{SeiGorSav-PRA-07}, we see from Table~\ref{tab_wprime} that the functional $W_\infty'[\rho]$ is much more seriously overestimated. We can reduce these errors by recalling that in the PC model for $W_\infty'[\rho]$  the coefficient $D$ 
of Eq.~\eqref{eq_WPC} was fixed by the condition that the PC value for the He atom be equal to the one obtained from the MGGA functional of Ref.~\cite{PerKurZupBla-PRL-99}. Now that we have exact values, it seems natural to change $D$ in order to make the PC model equal to the SCE result for the He atom. This gives $D=-0.028957$. The values for the other atoms obtained with the revised PC model are reported in Table~\ref{tab_revPC}: we see that the error is now substantially reduced.

%%%%%%%%%%%%%%%%%%%%%%%%%%%%
\begin{table}
\begin{tabular}{lllc}
\hline \hline
& SCE (H) & revPC (H) & error (mH)\\
\hline 
Li     &   $1.38$      &              $1.4066$       &            $26$ \\
Be     &  $2.59$       &             $2.579$       &            $11$ \\
B      &   $4.2$       &               $4.207$    &               $7$ \\
C      &   $6.3$       &               $6.43$    &               $130$ \\
Ne     &   $22$        &              $22.96$    &             $960$ \\
\hline \hline
\end{tabular}
\caption{Comparison of the values $W'_\infty[\rho]$ in Hartree atomic
units obtained with the SCE construction, and with the revised PC model of Sec.~\ref{sec_Wprime}. 
The absolute errors of the revised PC model are also reported.}
\label{tab_revPC}
\end{table}
%%%%%%%%%%%%%%%%%%%%%%%%%%%%%
\section{Revised ISI}
\label{sec_newISI}
In Refs.~\cite{SeiPerKur-PRA-00,SeiPerKur-PRL-00} an expression for $W_\lambda[\rho]$ that interpolates between the two limits of Eqs.~\eqref{wil} and \eqref{sil} has been proposed and tested using the PC approximation for the functionals $W_\infty[\rho]$ and $W'_{\infty}[\rho]$.
The interaction-strenght interpolation (ISI) formula for $W_\lambda[\rho]$ of Refs.~\cite{SeiPerKur-PRA-00,SeiPerKur-PRL-00}, however, contains a spurious term $\propto \lambda^{-1}$ in its $\lambda\to\infty$ expansion \cite{SeiPerKur-PRA-00}, which, as explained after Eq.~\eqref{eq_wprime}, has the wrong scaling behavior in the high-density limit. Here we propose a revised ISI functional which does not have this problem. 

Instead of modeling $W_\lambda[\rho]$, we use  the same ISI interpolation formula of Ref.~\cite{SeiPerKur-PRL-00} directly for the integral $E_{\rm xc}^\lambda[\rho]$,
\beq
E_{\rm xc}^\lambda[\rho]=\int_0^\lambda d\lambda' W_{\lambda'}[\rho],
\eeq
satisfying the exact $\lambda\to 0$ and $\lambda\to\infty$ asymptotic behaviors,
\beq
E_{\rm xc}^{\lambda,{\rm ISI}}[\rho]=a[\rho]\,\lambda +\frac{b[\rho]\,\lambda}{\sqrt{1+c[\rho]\,\lambda}+d[\rho]}.
\label{eq_ExclambdaISI}
\eeq
The four functionals $a[\rho]$, $b[\rho]$, $c[\rho]$ and $d[\rho]$ are determined by imposing the $\lambda\to 0$ expansion of  Eq.~\eqref{wil} and the $\lambda\to\infty$ expansion of Eq.~\eqref{sil}, and they are thus determined by the two weak-interaction limit functionals $E_x[\rho]$ and $E_c^{\rm GL2}[\rho]$ 
and the two strong-interaction limit functionals $W_{\infty}[\rho]$ and $W'_{\infty}[\rho]$,
\begin{eqnarray}
	a[\rho] & = & W_{\infty}[\rho]\\
	b[\rho] & = & -\frac{8\,E_c^{\rm GL2}[\rho] W'_{\infty}[\rho]^2}{(E_x[\rho]-W_{\infty}[\rho])^2} \\
	c[\rho] & = & \frac{16\,E_c^{\rm GL2}[\rho]^2 W'_{\infty}[\rho]^2}{(E_x[\rho]-W_{\infty}[\rho])^4} \\
	d[\rho] & = & -1-\frac{8\,E_c^{\rm GL2}[\rho] W'_{\infty}[\rho]^2}{(E_x[\rho]-W_{\infty}[\rho])^3}.
\end{eqnarray}
The final formula for the revised ISI functional is obtained by putting $\lambda=1$ in Eq.~\eqref{eq_ExclambdaISI},
\beq
E_{\rm xc}^{{\rm revISI}}[\rho]=a[\rho] +\frac{b[\rho]}{\sqrt{1+c[\rho]}+d[\rho]}.
\eeq
For the correlation energy of the neutral atoms considered here, this revised ISI gives essentially the same results of the original ISI of Ref.~\cite{SeiPerKur-PRL-00}.
\section{Conclusions and perspectives}
\label{sec_conc}
We have presented an exact treatment of the strong-interaction limit of density functional theory up to the second leading term, describing zero-point oscillations of strictly correlated electrons. We have evaluated numerically this zero-point contribution for small atoms, and we have used our results to improve a previous approximate functional for this term. A new interpolation formula for the exchange-correlation energy, satisfying more exact constraints, has been proposed, and will be tested elsewhere.

Besides the possibility of constructing an interpolation formula for $E_{\rm xc}[\rho]$, the two functionals $W_\infty[\rho]$ of Ref.~\cite{SeiGorSav-PRA-07} and $W_\infty'[\rho]$ evaluated in this work, are of valuable interest for the development of Kohn-Sham DFT. They are an example of exact implicit density functionals for systems in which the electron-electron repulsion largely dominates over the kinetic energy. They can be used to test properties of the exact exchange-correlation functional like the Lieb-Oxford bound \cite{LieOxf-IJQC-81,ChaHan-PRA-99}, and to test how approximate functionals perform in this limit \cite{StaScuTaoPer-PRB-04,JunGarAlvGod-PRA-04}. 

Several issues still need to be addressed and will be the object of future work. The main problem of the ISI functional is the lack of size consistency. In order to be size-consistent, the interpolation of Eq.~\eqref{eq_ExclambdaISI} should be done locally, using energy densities all defined in the same gauge. A first step in our future work will be the analysis of exact energy densities for the functionals $W_\infty[\rho]$ and $W'_\infty[\rho]$ (see also Ref.~\cite{GorSeiSav-PCCP-08}), and the construction of corresponding approximations. Another important problem is the development of a reliable algorithm to solve the SCE problem for a given non-spherical density. Other promising research lines are the study of the next leading term, which is of purely kinetic origin, and the construction of approximations to describe the effect of the spin state on the energy.

\section*{Acknowledgments}
We thank Mel Levy, Kieron Burke  and Andreas Savin for stimulating discussions. P.G.-G. was supported by ANR (07-BLAN-0272), and G.V. was supported by DOE  under Grant No.~DE-FG02-05ER46203.

\appendix
\section{Transformation of the Laplacian}
\label{app_T}
In order to write down the components of the metric tensor $G_{ik}$ of our local curvilinear coordinate transformation, we define the indices as follows: $\alpha,\beta,\gamma,..$ denote the cartesian components $1,2,3\equiv x,y,z$ of $\sv$, the indices $\mu,\nu,\sigma,\tau,...$ denote the normal-mode components $q_\mu$, and the latin indices $i,k,...$ denote general components, either $\alpha,..$ or $\mu,...$. 
We then have to distinguish three blocks in the metric tensor $G_{ik}$: $\alpha\beta$, $\mu\nu$, and $\alpha\mu$, 
\begin{eqnarray}
	G_{\beta\gamma}& = & g_{\beta\gamma}(\sv)-2\sum_{\mu=4}^{3N}q_{\mu}\frac{\partial^2 \Fv(\sv)}{\partial s_\gamma\partial s_{\beta}}\cdot \el^\mu(\sv)+ \nonumber \\
	&+ & \sum_{\mu,\nu=4}^{3N}\frac{\partial\el^\mu(\sv)}{\partial s_{\beta}}\cdot\frac{\partial \el^\nu(\sv)}{\partial s_{\gamma}}q_\mu q_\nu \label{eq_Galphabeta} \\
	G_{\beta\nu} & = & \sum_{\mu=4}^{3N}q_\mu \frac{\partial\el^\mu(\sv)}{\partial s_{\beta}}\cdot \el^\nu(\sv) \\
	G_{\mu\nu} & = & \delta_{\mu\nu},
\end{eqnarray}
where in Eq.~\eqref{eq_Galphabeta} we have defined the $3\times 3$ metric tensor $g_{\alpha\beta}(\sv)$ which only concerns the coordinates $s_1,s_2,s_3$,
\beq
g_{\alpha\beta}(\sv)=\frac{\partial \Fv(\sv)}{\partial s_\alpha}\cdot \frac{\partial \Fv(\sv)}{\partial s_\beta}.
\eeq
When $\lambda\to\infty$, our wavefunction is zero everywhere except very close to $\itOm_0$, i.e., for very small $q_\mu\propto \lambda^{-1/4}$. Introducing the scaled coordinates $u_\mu=\lambda^{1/4}q_\mu$, we see that the metric tensor $G_{ik}$ has the $\lambda$-dependence
\beq
G_{ik}=G_{ik}^{(0)}+\frac{1}{\lambda^{1/4}}\sum_{\mu=4}^{3N}u_\mu \Delta^\mu_{ik}+\frac{1}{\lambda^{1/2}}\sum_{\mu,\nu=4}^{3N}u_\mu u_\nu Z^{\mu\nu}_{ik},
\eeq
where $\Delta^\mu$ and $Z^{\mu\nu}$ are tensors of elements
\begin{eqnarray}
	\Delta^\mu_{\gamma\beta} & = & -2\frac{\partial^2 \Fv(\sv)}{\partial s_\gamma\partial s_{\beta}}\cdot \el^\mu(\sv) \\
	\Delta^\mu_{\beta\nu} & = & \frac{\partial\el^\mu(\sv)}{\partial s_{\beta}}\cdot \el^\nu(\sv)\\
	\Delta^\mu_{\nu\tau} & = & 0,
\end{eqnarray}
and
\begin{eqnarray}
	Z^{\mu\nu}_{\beta\gamma}& = & \frac{\partial\el^\mu(\sv)}{\partial s_{\beta}}\cdot\frac{\partial \el^\nu(\sv)}{\partial s_{\gamma}} \\
	Z^{\mu\nu}_{\beta\tau} & = & 0 \\
	Z^{\mu\nu}_{\tau\sigma} & = & 0, 
\end{eqnarray}
and $G^{(0)}_{ik}$ has elements $G^{(0)}_{\alpha\beta}=g_{\alpha\beta}$, $G^{(0)}_{\mu\nu}=\delta_{\mu\nu}$ and all the off-diagonal components equal to zero.
In order to compute the large-$\lambda$ expansion of Eq.~\eqref{eq_lapl}, we have to expand the determinant $G$, and the components $G^{ik}$ of the inverse metric tensor. Using standard formulas, we obtain
\beq
\sqrt{G}  =  \sqrt{g}\left(1+\frac{1}{2\lambda^{1/4}}\sum_{\mu=4}^{3N}u_\mu\sum_{\alpha\beta}g^{\alpha \beta}\Delta^\mu_{\alpha\beta}\right)+O\left(\lambda^{-1/2}\right),
\label{Jac}
\eeq
where $g$ is the determinant of $g_{\alpha\beta}$, and $g^{\alpha\beta}$ are the components of its inverse. The tensor ${\bf G}^{-1}$ of components $G^{ik}$ has the large-$\lambda$ expansion, up to orders $\lambda^{-1/2}$,
\beq
{\bf G}^{-1}={\bf G^{(0)}}^{-1}-\frac{1}{\lambda^{1/4}}\sum_{\mu=4}^{3N}u_\mu{\bf G^{(0)}}^{-1}{\bf \Delta}^\mu {\bf G^{(0)}}^{-1}. 
\eeq
Inserting these expansions into Eq.~\eqref{eq_lapl} we obtain Eqs.~\eqref{eq_T1} and \eqref{eq_T2}
with 
\beq
X^\mu(\sv)=\frac12\sum_{\alpha\beta}g^{\alpha\beta}(\sv)\Delta^\mu_{\alpha\beta}(\sv).
\eeq

Finally, the Jacobian of our change of coordinates is simply equal to $\sqrt{G}$
of Eq.~\eqref{Jac}.

\section{Analytic example}
\label{app_ana}

As an illustration, we consider a system of two electrons in 1D space (i.e., on the $x$-axis)
with a given ground-state density
$\rho(x)$,
\beq
\int_{-\infty}^{\infty}dx\,\rho(x)=2.
\eeq
In this case, Eq.~\eqref{cofs} reads $x_2=f_2(x_1)$, with the single co-motion function $f_2(s)\equiv f(s)$ which,
according to Ref. \cite{SeiGorSav-PRA-07}, obeys the differential equation $\rho(f(s))f'(s)=\rho(s)$.
For the Lorentzian density, $f(s)$ is found analytically, 
\beq
\rho(x)=\frac2{\pi}\frac1{1+x^2}\qquad\Rightarrow\qquad f(s)=-\frac1s.
\label{rhoL}
\eeq
In this case, the SCE external potential, fixed by the conditions $\frac{d}{dx}v_{\rm SCE}(x)=\mbox{sgn}(x)|x-f(x)|^{-2}$ and 
$v_{\rm SCE}(x)\to0$ for $x\to\pm\infty$, is given by
\beq
v_{\rm SCE}(x)=\Big|\arctan(x)-\frac{x}{1+x^2}\Big|-\frac{\pi}2.
\eeq
In terms of $\Fv(s)\equiv(s,f(s))$, Eq.~\eqref{Om0} now yields a 1D subset of $\itOm={\sf R}^2$,
\beq
\itOm_0=\{\Fv(s)|s\in{\sf R}\}\subset\itOm.
\eeq
In the example \eqref{rhoL}, $\itOm_0$ is given by the two branches of the hyperbola $x_2=f(x_1)\equiv-\frac1{x_1}$
in the $x_1x_2$-plane $\itOm$. In the following, we focus on the branch $\itOm_0^+$ with $x_1>0$ and $x_2<0$,
$\itOm_0^+=\{\Fv(s)|s\in{\sf R}^+\}$.

The asymptotic potential-energy function, cf.~Eq.~\eqref{Epot},
\beq
E_{\rm pot}(\Xv)=\frac1{x_1-x_2}+v_{\rm SCE}(x_1)+v_{\rm SCE}(x_2),
\label{eq_Epotana}
\eeq
assumes its highly degenerate minimum for all $\Xv\in\itOm_0$. Consequently,
the first partial derivatives,
\begin{eqnarray}
\frac{\partial E_{\rm pot}(\Xv)}{\partial x_1}&=&-\frac1{(x_1-x_2)^2}+\frac{x_1^2}{(1+x_1^2)^2},\nonumber\\
\frac{\partial E_{\rm pot}(\Xv)}{\partial x_2}&=&+\frac1{(x_1-x_2)^2}-\frac{x_2^2}{(1+x_2^2)^2},
\end{eqnarray}
are vanishing for $\Xv=\Fv(s)$ when the Hessian matrix of $E_{\rm pot}(\Xv)$ becomes
\beq
M(s)=\frac{2s}{(1+s^2)^3}\left(\begin{array}{cc} 1 & -s^2 \\ -s^2 & s^4 \end{array}\right).
\eeq
It has the two eigenvalues
\beq
\omega_1(s)^2=0,\qquad\omega_2(s)^2=\frac{2s}{(1+s^2)^3}(1+s^4)>0.
\eeq
The corresponding normalized eigenvectors are
\beq
\el^1(s)=\frac1{\sqrt{1+s^4}}\Big(\begin{array}{c} s^2 \\ 1    \end{array}\Big),\qquad
\el^2(s)=\frac1{\sqrt{1+s^4}}\Big(\begin{array}{c} 1   \\ -s^2 \end{array}\Big).
\eeq
While $\el^1(s)$ is tangential, $\el^2(s)$ is orthogonal to $\itOm_0^+$ at $\Fv(s)\in\itOm_0^+$ and
generally given by
\beq
\el^2(s)=\frac1{\sqrt{1+f'(s)^2}}\Big(\begin{array}{c} f'(s) \\ -1 \end{array}\Big)\equiv\el(s).
\label{w2gen}
\eeq

%\newpage
For a point $\Xv=(x_1,x_2)\in\itOm_\eps$, close to $\itOm_0^+$, the curvilinear coordinates
$(s,q)$ are defined by Eq.~\eqref{qr},
\beq
\Xv=\Fv(s)+\el(s)\,q
\label{xxsq}
\eeq
where $s$ is fixed by the condition that the vector $\el(s)$ in the $x_1x_2$-plane is orthogonal
to $\itOm_0^+$ at $\Fv(s)\in\itOm_0^+$.

In terms of the partial derivatives of Eq.~\eqref{xxsq}, the metric tensor is given by the $(2\times2)$-matrix
\beq 
(G_{\mu\nu})=\left(\begin{array}{ccc}
\displaystyle\frac{\partial\Xv}{\partial s}\cdot\frac{\partial\Xv}{\partial s} & &
\displaystyle\frac{\partial\Xv}{\partial s}\cdot\frac{\partial\Xv}{\partial q} \\
\displaystyle\frac{\partial\Xv}{\partial q}\cdot\frac{\partial\Xv}{\partial s} & &
\displaystyle\frac{\partial\Xv}{\partial q}\cdot\frac{\partial\Xv}{\partial q} \end{array}\right).
\eeq
Using Eqs.~\eqref{w2gen} and \eqref{xxsq}, we obtain
\begin{eqnarray}
\frac{\partial\Xv}{\partial s} &=& \Fv'(s)+q\el'(s)\nonumber\\
                               &=& \Big(\begin{array}{c} 1 \\ f'(s) \end{array}\Big)
                           \Big[1+q\frac{f''(s)}{[1+f'(s)^2]^{3/2}}\Big],\nonumber\\
\frac{\partial\Xv}{\partial q} &=& \el(s)
\end{eqnarray}
and thus
\beq
(G_{\mu\nu})=\left(\begin{array}{cc} g(s,q) & 0 \\ 0 & 1 \end{array}\right)
\eeq
where $g(s,q)=J(s,q)^2$, with the Jacobian
\beq
J(s,q)=\Big[1+q\frac{f''(s)}{[1+f'(s)^2]^{3/2}}\Big]\sqrt{1+f'(s)^2}.
\eeq

In the particular case of the density \eqref{rhoL}, we have
\beq
J(s,q)=\frac{\sqrt{1+s^4}}{s^2}-q\frac{2s}{1+s^4},
\eeq
and the coefficients of Eq.~\eqref{sil} are given by
\begin{eqnarray}
W_\infty[\rho]+U[\rho]&\equiv&2\int_0^\infty ds\frac{\rho(s)}2\,\frac1{s-f(s)}\nonumber\\
&=&\frac1\pi\int_0^\infty\frac{ds\,2s}{(1+s^2)^2}=\frac1\pi\nonumber\\
&=&0.31831.
\end{eqnarray}
\begin{eqnarray}
W'_\infty[\rho]&\equiv&2\int_0^\infty ds\frac{\rho(s)}2\,\omega_2(s)\nonumber\\
&=&\frac2\pi\int_0^\infty\frac{ds}{(1+s^2)^2}\sqrt{2s\,\frac{1+s^4}{1+s^2}}\nonumber\\
&=&0.633902.
\end{eqnarray}

%%%REFERENCES

\end{document}